# SIMULATING POYNTING FLUX ACCELERATION IN THE LABORATORY WITH COLLIDING LASER PULSES

EDISON LIANG

*Rice University, Hsouton, TX 77005-1892 USA*

**Abstract** We review recent PIC simulation results which show that double-sided irradiation of a thin over-dense plasma slab with ultra-intense laser pulses from both sides can lead to sustained comoving Poynting flux acceleration of electrons to energies much higher than the conventional ponderomotive limit. The result is a robust power-law electron momentum spectrum similar to astrophysical sources. We discuss future ultra-intense laser experiments that may be used to simulate astrophysical particle acceleration.

**Keywords:** electron acceleration, laser-plasma interaction, laboratory astrophysics

## 1. Introduction

Most high energy astrophysical sources (pulsars, blazars, gamma-ray bursts, supernova remnants) emit a simple power-law spectrum in the x-gamma-ray range. The most common observed photon index lies in the range 2-3, which translates into an electron momentum index of 3 – 5 for optically thin radiation (Rybicki and Lightman 1979). The most popular current models for astrophysical particle acceleration are shock acceleration (first-order Fermi), diffusive wave acceleration, and Poynting flux acceleration by large-scale electromagnetic fields. In earlier work (Liang et al 2003) we demonstrated that Poynting flux acceleration driven by electromagnetic-dominated outflows (Liang et al 2003) naturally produces robust power-law relativistic electron spectra. Poynting flux acceleration of e+e- plasmas is especially relevant to gamma-ray bursts and pulsar winds. It is therefore highly desirable to study particle acceleration in the laboratory that may mimic or at least shed new light on Poynting flux acceleration in astrophysics.

Recent advances in ultra-intense short-pulse lasers (ULs) (Mourou et al 1998, Ditmire 2003) open up new frontiers on particle acceleration by ultra-strong electromagnetic (EM) fields in plasmas (Lontano et al 2002). However, most conventional laser acceleration schemes (e.g. laser wakefield accelerator, plasma wakefield accelerator, plasma beat-wave accelerator, free wave accelerator, see Esarey et al 1996, Sprangle et al 1990; Malka 2002, Pukhov et al 1997, Tajima and Dawson 1979, Hussein et al 1992, Kawata et al 1991, Woodworth et al 1996) involve the propagation of lasers in an underdense plasma ($\omega_{pe}=(4\pi ne^2/m_e)^{1/2}<\omega_o=2\pi c/\lambda$, $\lambda$=laser wavelength, n=electron density). In such schemes the acceleration gradient (energy gain/distance) (Esarey et al 1996, Malka 2002) and energetic particle beam intensity are limited by the underdense requirement. They also do not produce a power-law electron spectrum.

Here we review PIC simulation results of a radically different concept: comoving acceleration of overdense ($\omega_{pe} > \omega_o$) plasmas using colliding UL pulses. In this case the acceleration gradient and particle beam intensity are not limited by the underdensity condition.  This colliding laser pulses accelerator (CLPA) concept may have important applications to laboratory astrophysics since CLPA naturally produces a power-law electron spectrum, similar to the high energy spectra of observed astrophysical sources.  Most other laser acceleration schemes produce either exponential or quasi-monoenergetic electron momentum distributions.

## 2. Colliding Laser Pulses Accelerator

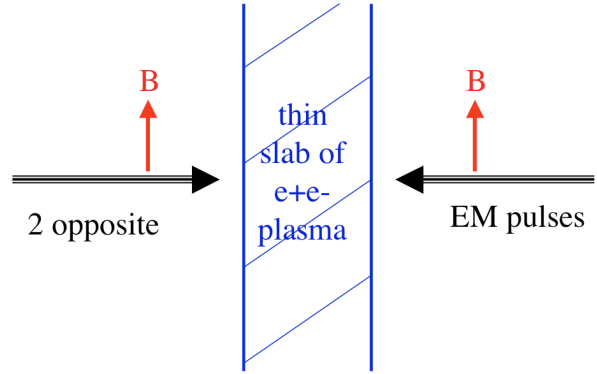

Fig.1. Schematic diagram showing the CLPA concept.

Fig.1 shows the basic idea of CLPA.  Two linear polarized intense laser pulses with aligned **B** vectors irradiate a thin overdense plasma slab from opposite sides.  They compress the slab until it becomes thinner than 2 relativistic skin depths.  At that point the laser pulses "tunnel through" and capture the surface electrons as they reemerge at the far side of the slab.  Due to plasma loading the laser pulses slow down and stay in phase with the fastest particles, and accelerate them continuously with self-induced comoving **J x B** forces.  Fig. 2 shows the PIC simulation of two linearly polarized plane half-cycle EM pulses with parallel **B**, irradiating a thin e+e- slab from opposite sides (thickness=$\lambda/2$, initial density $n_o=15n_{cr}$(critical density)).  Cases with nonparallel **B** are more complex and are still under investigation.  Each incident pulse compresses and accelerates the plasma inward (Fig.1a), reaching a terminal Lorentz factor of $\gamma_{max} \sim (\Omega_e/\omega_{pe})^2 \sim 40$. Only ~10% of the incident EM amplitudes is reflected because the laser reflection front is propagating inward relativistically (Kruer et al 1975).  As the relativistic skin depths from both sides start to merge (Fig.1b), the two UL pulses interpenetrate and tunnel through the plasma, despite $\omega_{pe} > <\gamma>^{1/2} \omega_o$.  Such transmission of EM waves through an overdense plasma could not be achieved using a single UL pulse, because there the upstream plasma is snowplowed by the laser pressure indefinitely. As the transmitted UL pulses reemerge from the plasma, they induce new drift currents **J** at the *trailing* edge of the pulses (Fig.1c), with opposite signs to the initial currents (Fig.1b), so that the new **J x B** forces pull the surface plasmas outward.  We emphasize that the plasma loading which slows the transmitted UL pulses plays a crucial role in sustaining this comoving acceleration.  For a given $\Omega_e/\omega_{pe}$ the higher the plasma density, the more sustained the comoving acceleration, and a larger fraction of the plasma slab is accelerated. This unique feature distinguishes this overdense acceleration scheme from other underdense schemes. As slower particles gradually fall behind the UL pulses, the plasma loading of the UL pulses decreases with time. This leads to continuous acceleration of both the UL pulses and the dwindling population of trapped fast particles . The phase space evolution (Fig.1d) of this colliding laser pulses accelerator (CLPA) resembles that of the DRPA discovered earlier (Liang et al 2003, 2004, Nishimura et al 2004).

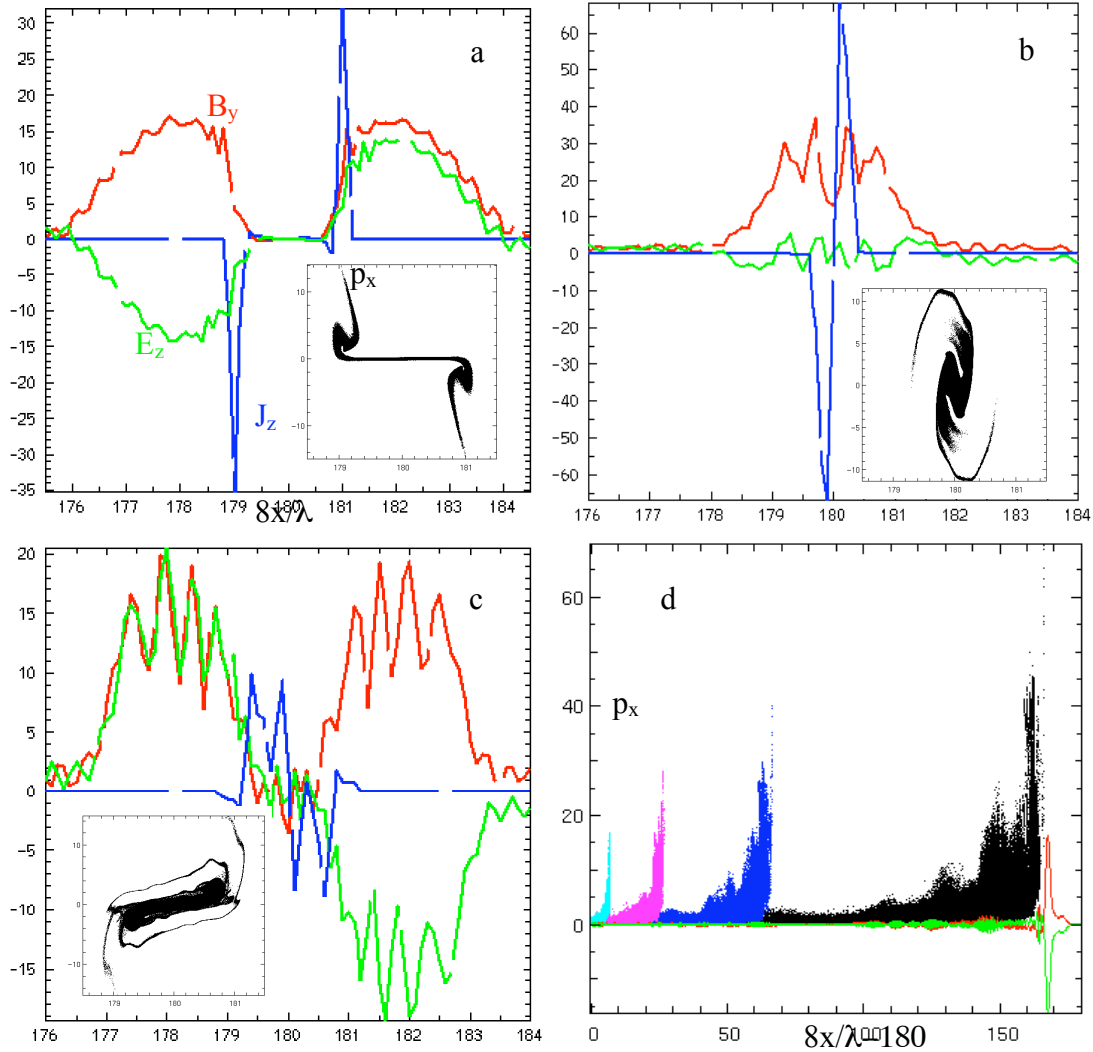

Fig.2. Evolution of two linearly polarized plane EM pulses ($I(\lambda/\mu m)^2=10^{21}W/cm^2$, $c\tau=\lambda/2$) irradiating an overdense e+e- plasma ($n_o=15n_{cr}$, thickness = $\lambda/2$, kT=2.6keV) from opposite sides. We plot magnetic field $B_y$(medium), electric field $E_z$(light), current density $J_z$(dark) and $p_x/mc$ vs. x (inset) at $t\omega_o/2\pi$ = (a)1.25, (b)1.5, (c)1.75; (d) Snapshots of $p_x/m_ec$ vs. x (dots) for the right-moving pulse at $t\omega_o/2\pi$=2.5(black), 5(red), 10(blue), 22.5(green) showing power law growth of $\gamma_{max}\sim t^{0.45}$. We also show the profiles of $B_y$(medium), $E_z$(light) at $t\omega_o/2\pi$=22.5 (from Liang 2006).

## 3. Acceleration by Colliding Gaussian Laser Pulse Trains

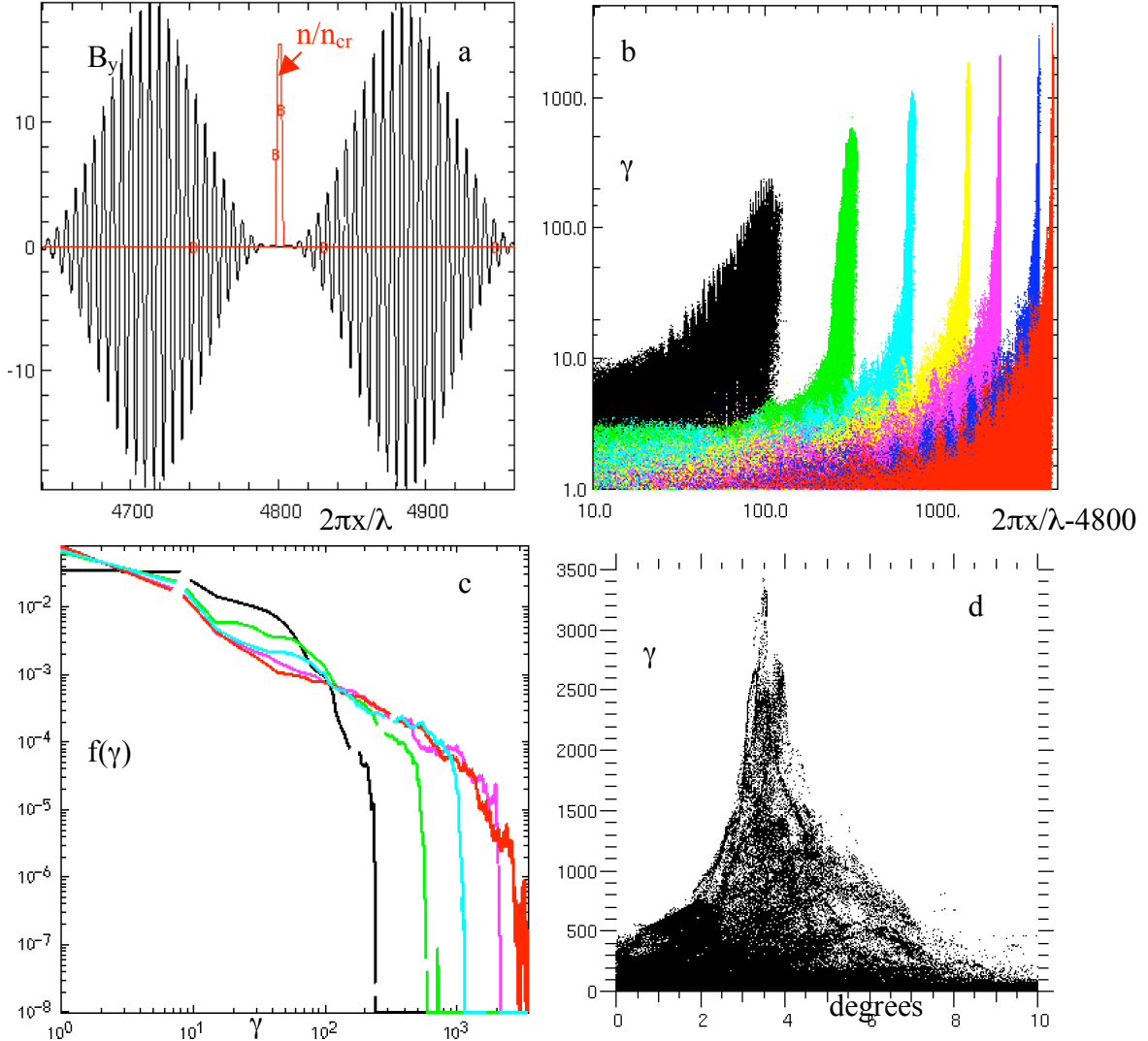

Fig.3. Results of two Gaussian pulse trains ($\lambda=1\mu m$, $I=10^{21}W/cm^2$, $c\tau=85fs$) irradiating a e+e- plasma ($n_o=9n_{cr}$, thickness = $2\lambda/\pi$, kT=2.6keV). (a) early $B_y$ and $n_o/n_{cr}$ (B) profiles at $t\omega_o=0$; (b) time-lapse evolution of $\log(p_x/m_ec)$ vs. logx for the right-moving pulse at $t\omega_o=$ (left to right) 180, 400, 800, 1600, 2400, 4000, 4800 showing power-law growth of $\gamma_{max}\sim t^{0.8}$; (c) evolution of electron energy distribution $f(\gamma)$ vs. $\gamma$ showing the build-up of power-law below $\gamma_{max}$ with slope $\sim -1$: $t\omega_o=$ (left to right) 180, 400, 800, 2400, 4800. (Slope =–1 means equal number of particles per decade of energy), (d) plot of $\gamma$ vs. $\theta$ (=$|p_z|/|p_x|$) in degrees at $t\omega_o=4800$, showing strong energy-angle selectivity and narrow beaming of the most energetic particles (from Liang 2006)

Fig.3 shows the results of irradiating an overdense e+e- slab using more realistic Gaussian pulse trains ($\lambda=1\mu m$, pulse length $\tau=85fs$, $I_{peak}=10^{21}Wcm^{-2}$). We see that $\gamma_{max}$ increases rapidly to 2200 by 1.28ps and 3500 by 2.56ps, far exceeding the ponderomotive limit $a_o^2/2$ (~360). The maximum Lorentz factor increases with time according to $\gamma_{max}(t)\sim e\int E(t)dt/mc$. E(t) is the UL electric field comoving with the highest energy particles. E(t) decreases with time due to EM energy transfer to the particles, plus slow dephasing of particles from the UL pulse peak. This leads to $\gamma_{max}$ growth slower than linear and $\gamma_{max} \sim t^{0.8}$ (Fig.2b). In practice, $\gamma_{max}$ will be limited by the diameter D of the laser focal spot, since particles drift transversely out of the laser field after $t \sim D/c$. The maximum energy of any comoving acceleration is thus $< eE_oD=6GeV(I/10^{21}Wcm^{-2})^{1/2}(D/100\mu m)$. The asymptotic momentum distribution forms a power-law with slope $\sim -1$ (Fig.2d) below $\gamma_{max}$, distinct from the exponential distribution of ponderomotive heating (Kruer et al 1985, Wilks et al 1992, Gahn et al 1999, Wang et al 2001, Sheng et al 2004). A quasi-power-law momentum distribution is formed below $\gamma_{max}$ since there is no other preferred energy scale below $\gamma_{max}$, and the particles have random phases with respect to the EM field profile.

## 4. Proposed Laser Experiment

An experimental demonstration of the CLPA will require a dense and intense e+e- source. (Cowan et al 1999, 2000) demonstrated that such an e+e- source can be achieved by using a PW laser striking a gold foil. Theoretical works (Liang et al 1998, Shen et al 2001) suggest that e+e- densities $>10^{22}cm^{-3}$ may be achievable with sufficient laser fluence. Such a high density e+e- jet can be slit-collimated to produce a $\sim$ micron thick e+e- slab, followed by 2-sided irradiation with opposite UL pulses. As an example, consider UL pulses with $\tau=80fs$ and intensity=$10^{19}Wcm^{-2}$. We need focal spot diameter D>600 $\mu$m for the pairs to remain inside the beam for >1ps. This translates into ~1KJ energy per UL pulse. Such high-energy UL's are currently under construction at many sites (Ditmire 2003). Fig.3 shows the artist conception of such an experimental setup.

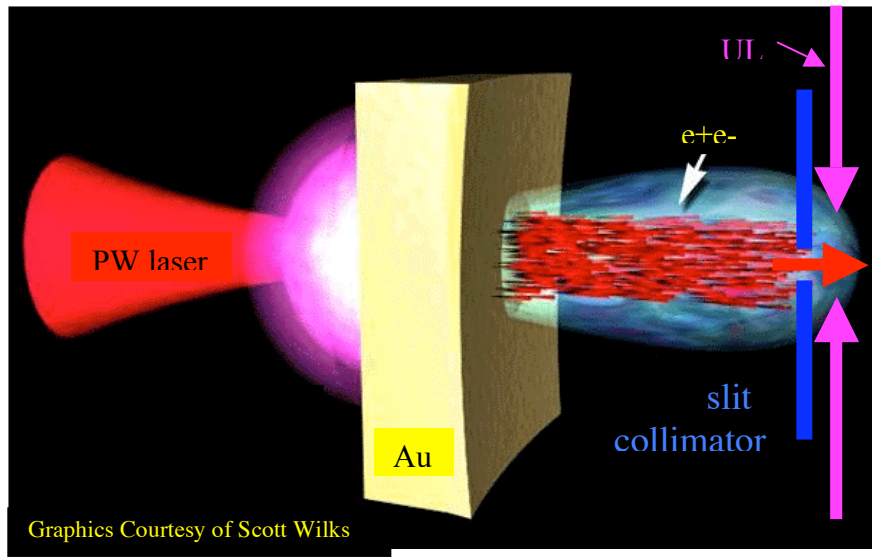

Fig.4. Conceptual experimental setup for the demonstration of the CLPA mechanism using three PW lasers.

We have also performed simulations of CLPA using electron-ion plasmas. Results so far suggest that as long the e-ion slab is sufficiently thin and laser pulses sufficiently intense, so that the electrons can be compressed to less than two relativistic skin depths before the lasers are reflected, the electrons are accelerated by the reemerging pulses similar to the e+e- case. However the ions lag behind the electrons due to their inertia and are accelerated only by the charge-separation electric field. The late-time partition between electron and ion energies depends on the plasma density and laser intensities. Note that CLPA is insensitive to the relative phases of the two pulses. If one pulse arrives first it simply pushes the plasma toward the other pulse until it hits. Then both pulses compress the slab together with the same final results.

## Acknowledgement

EL was partially supported by NASA NAG5-9223, LLNL B537641 and NSF AST-0406882. He thanks Scott Wilks for help with running ZOHAR and the graphics of Fig.4, and Bruce Langdon for providing the ZOHAR code.

## References

T.E. Cowan et al. 1999 *Laser Part. Beams* **17**, 773; ibid, 2000 *Phys. Rev. Lett.* **84**, 903.
T. Ditmire, ed. 2003 *SAUUL Report*, (UT Austin).
E. Esarey, P. Sprangle, J. Krall, A. Ting, 1996 IEEE Trans. Plasma Sci. **24**, 252.
C. Gahn et al 1999 Phys. Rev. Lett. 83, 4772.
M.S. Hussein, M.P. Pato, A.K. Kerman 1992 *Phys. Rev. A* **46**, 3562.
M.S. Hussein, M.P. Pato 1991 *Phys. Rev. Lett.* **68**, 1992.
S. Kawata, T. Maruyama, H. Watanabe, I. Takahashi 1991 *Phys. Rev. Lett.* **66**, 2072.
W.L. Kruer, E.J. Valeo, K.G. Estabrook 1975 *Phys. Rev. Lett.* **35**, 1076.
W.L. Kruer, K.G. Estabrook 1985 *Phys. Fluids* **28**, 430.
E. Liang, K. Nishimura, H. Li, S.P. Gary,2003 *Phys. Rev. Lett.* **90**, 085001.
E. Liang, K. Nishimura 2004 *Phys. Rev. Lett.* **92**, 175005.
E. Liang 2006 *Phys. Plasmas* in press.
E.P. Liang, S.C. Wilks, M. Tabak 1998 *Phys. Rev. Lett.* **81**, 4887.
L Lontano et al, eds. 2002 *Superstrong Fields in Plasmas*, AIP Conf. Proc. No. **611** (AIP, NY).
V. Malka 2002 in *AIP Conf. Proc.* No. **611**, p.303, ed. M. Lontano et al. (AIP, NY)
G.A. Mourou, C.P.J. Barty, M.D. Perry 1998 *Phys. Today* **51**(1), 22.
K. Nishimura, E. Liang 2004 *Phys. Plasmas* **11** (10).
A. Pukhov, J. Meyer-ter-Vehn 1997 *Phys. Rev. Lett.* **79**, 2686.
G. Rybicki and A.P. Lightman 1979 *Radiative Processes in Astrophysics* (Wiley, NY).
B. Shen, J. Meyer-ter-Vehn 2001 *Phys. Rev. E* **65**, 016405.
Z.M. Sheng, K. Mima, J. Zhang, J. Meyer-ter-Vehn 2004 *Phys. Rev. E* **69**, 016407.
P. Sprangle, E. Esary, A. Ting 1990 *Phys. Rev. Lett.* **64**, 2011.
T. Tajima and J.M. Dawson 1979 *Phys. Rev. Lett.* **43**, 267.
P.X. Wang et al. 2001 App. Phys. Lett. 78, 2253.
S.C. Wilks, W.L. Kruer, M. Tabak, A.B. Langdon 1992 *Phys. Rev. Lett.* **69**, 1383.
J.G. Woodworth, M.N. Kreisler, A.K. Kerman 1996 *The Future of Accelerator Phys.* p.378, ed. T. Tajima (AIP, NY).